\documentclass[10pt, conference]{IEEEtran}
\newcommand{\subsectionRemoveSpaceUnderSection}{0pt}
\newcommand{\subsectionRemoveSpace}{0pt}
\newcommand{\sectionRemoveSpace}{0pt}

\usepackage[T1]{fontenc}
\usepackage{tikz, tikz-3dplot}
\usetikzlibrary{positioning, patterns}
\usetikzlibrary{arrows,automata, shapes, petri, matrix}
\usetikzlibrary{arrows.meta, shapes.callouts, math}
\usetikzlibrary{calc,decorations.pathreplacing}
\usepackage{tikz-timing}
\usetikztiminglibrary[new={char=Q,reset char=R}]{counters}
\usepackage{pgfplots}
\usetikzlibrary{fadings}

\tikzset{router/.style={circle,draw,fill=gray!60,inner sep=0pt,minimum size=5pt}}
\tikzset{desc/.style={font = \scriptsize}}
\tikzset{cross/.style={cross out, draw, 
		minimum size=2*(#1-\pgflinewidth), 
		inner sep=0pt, outer sep=0pt}}

\definecolor{col1}{RGB}{27,158,119}
\definecolor{col2}{RGB}{217,95,2}
\definecolor{col3}{RGB}{117,112,179}

\definecolor{cFilter}{RGB}{217,95,2}
\definecolor{cInput}{RGB}{117,112,179}
\definecolor{cOutput}{RGB}{27,158,119}

\definecolor{Cyan}{rgb}{0.0, 1.0, 1.0}
\definecolor{Magenta}{rgb}{1.0, 0.0, 1.0}
\definecolor{ForestGreen}{rgb}{0.0, 0.27, 0.13}
\definecolor{OrangeRed}{rgb}{1.0,0.27,0.0}
\definecolor{OliveGreen}{rgb}{.332,.42,.185}
\definecolor{Goldenrod}{rgb}{.855,.648,.125}
\definecolor{NavyBlue}{rgb}{.0,.0,.5}
\definecolor{Plum}{rgb}{.868,.628,.868}
\definecolor{MidnightBlue}{rgb}{.10,0.10,0.44}
\definecolor{BrickRed}{rgb}{0.80,0.25,0.33}

\tikzstyle{myMarkerPlot}=[mark size=2.5pt, thick, densely dotted, mark options={solid,draw=black}]
\usepackage{pdfpages}

\usepackage[format=plain,textfont=it]{caption}
\usepackage{subcaption}

\usepackage{amsmath}
\usepackage{units}
\usepackage{tabularx}
\newcolumntype{Z}{>{\raggedright\let\newline\\\arraybackslash}X}
\usepackage{multirow}
\usepackage{rotating}

\usepackage{booktabs} % For formal tables

\definecolor{MyLightGreen}{HTML}{66CC33}
\definecolor{MyPurple}{HTML}{9900CC}
\definecolor{MyLightBlue}{HTML}{0066FF}
\definecolor{tablegray}{rgb}{0.95,.95,.95}
\usepackage{csvsimple}
\usepackage{multicol}
\usepackage{enumitem}
\usepackage{colortbl}
\usepackage{lscape}
\usepackage{adjustbox}
\usepackage{textcomp}
\setlist[description]{leftmargin=\parindent}

\usepackage{mathtools}
\usepackage{comment}

\newcommand{\squishlist}{
 \begin{list}{$\bullet$}
  { \setlength{\itemsep}{0pt}
     \setlength{\parsep}{3pt}
     \setlength{\topsep}{3pt}
     \setlength{\partopsep}{0pt}
     \setlength{\leftmargin}{1.5em}
     \setlength{\labelwidth}{1em}
     \setlength{\labelsep}{0.5em} } }

\newcommand{\squishlisttwo}{
 \begin{list}{$\bullet$}
  { \setlength{\itemsep}{0pt}
     \setlength{\parsep}{0pt}
    \setlength{\topsep}{0pt}
    \setlength{\partopsep}{0pt}
    \setlength{\leftmargin}{2em}
    \setlength{\labelwidth}{1.5em}
    \setlength{\labelsep}{0.5em} } }

\newcommand{\squishend}{
  \end{list}  }

\begin{document}
\bstctlcite{IEEEexample:BSTcontrol}

%%% Authors
% Joseph, Jan Moritz; RWTH Aachen University, Germany; joseph@ice.rwth-aachen.de
% Samajdar, Anand; Georgia Institute of Technology, Atlanta, GA; anandsamajdar@gatech.edu
% Lingjun Zhu; Georgia Institute of Technology, Atlanta, GA; lingjun@gatech.edu
% Leupers, Rainer; RWTH Aachen University, Germany; leupers@ice.rwth-aachen.de
% Lim, Sung-Kyn; Georgia Institute of Technology, Atlanta, GA; limsk@ece.gatech.edu
% Pionteck, Thilo; Otto-von-Guericke Universitaet Magdeburg, Germany; thilo.pionteck@ovgu.de
% Krishna, Tushar; Georgia Institute of Technology, Atlanta, GA; tushar@ece.gatech.edu

\title{Architecture, Dataflow and Physical Design Im\-pli\-cations of 3D-ICs for DNN-Accelerators}%\vspace{-1.7cm}}

\author{
	\IEEEauthorblockN{{Jan Moritz Joseph}\IEEEauthorrefmark{1}\IEEEauthorrefmark{2}, {Ananda Samajdar}\IEEEauthorrefmark{2}, {Lingjun Zhu}\IEEEauthorrefmark{2}, {Rainer Leupers}\IEEEauthorrefmark{1}, \\{Sung-Kyu Lim}\IEEEauthorrefmark{2}, {Thilo Pionteck}\IEEEauthorrefmark{3}, {Tushar Krishna}\IEEEauthorrefmark{2}}\\
	\IEEEauthorblockA{
	\IEEEauthorrefmark{1}RWTH Aachen University, Germany\\
			Email: \{joseph, leupers\}@ice.rwth-aachen.de\vspace{.7ex}}
	\IEEEauthorblockA{\IEEEauthorrefmark{3}Otto-von-Guericke-University Magdeburg, Germany\\
				Email: \{thilo.pionteck\}@ovgu.de\vspace{.7ex}}
	\IEEEauthorblockA{\IEEEauthorrefmark{2}Georgia Institute of Technology, Atlanta, GA\\
			Email: \{anandsamajdar,limsk, tushar\}@ece.gatech.edu}
}

% The default list of authors is too long for headers}
%\renewcommand{\shortauthors}{J. M. Joseph et al.}
\maketitle

\begin{abstract}
The everlasting demand for higher computing power for deep neural networks (DNNs) drives the development of parallel computing architectures. 3D integration, in which chips are integrated and connected vertically, can further increase performance because it introduces another level of spatial parallelism. Therefore, we analyze dataflows, performance, area, power and temperature of such 3D-DNN-accelerators. Monolithic and TSV-based stacked 3D-ICs are compared against 2D-ICs. We identify workload properties and architectural parameters for efficient 3D-ICs and achieve up to 9.14x speedup of 3D vs. 2D. We discuss area-performance trade-offs. We demonstrate applicability as the 3D-IC draws similar power as 2D-ICs and is not thermal limited.
\end{abstract}

%\keywords{ACM proceedings, \LaTeX, text tagging}

%\maketitle

%\todo{Requirements: 100 words abstract, 6 pages incl. references}

%\vspace{-10pt}
\section{Introduction}
%\vspace{-2pt}

Deep neural network (DNN) inference demands high computation and is inherently parallel. Recently, the popularity of DNNs has given rise to specialized accelerators \cite{eyeriss, tpu-isca}. Almost all DNN accelerators are matrix multiplication machines, since computation of DNNs follows this linear algebra motif, e.g., convolutions in CNNs (Convolution Neural Networks) or LSTM/GRU layers (Long Short Term Memory / Gated Recurrent Units) in Recurrent Neural Networks (RNNs).

The planar nature of modern accelerators allows two-dimensional parallelism. Executing operations with higher dimensions--e.g., convolutions with a batch size of one is a six-dimensional operation--requires unrolling the operands in 2D matrices, mapped along the two spatial dimensions and in time. This two-dimensional mapping limits the parallelism; Even with an infinitely large array, some operations must be executed sequentially. 

\begin{figure}
	\centering
	\scalebox{.9}{%\documentclass[border=2mm]{standalone}
%\usepackage[dvipsnames]{xcolor}
%\usepackage{pgfplots}
%\pgfplotsset{compat=1.8}
%\usepackage{amssymb}
%\usepackage{units}
%\usetikzlibrary{decorations.pathreplacing, decorations, positioning, calc, patterns, arrows, math, shadings}
%\input{defs.tex}
%
%\begin{document}
\begin{tikzpicture}[scale = .3, yscale = -1]
\label{fig:3d-array}
\newcommand{\drawLayer}[1]{
	%layer
	\draw[col1, fill = col1!50, opacity = .8] (0,#1, 0) -- (0,#1,6) --(10,#1,6) -- (10,#1,0) -- cycle;% opacity= 0.1
	%links
	\foreach \x in {1,2, ..., 8}{
		\foreach \y in {1, 2, ..., 5}{
			\draw[fill = black!10] (\x, #1, \y) -- (\x+1, #1, \y);
		}
	}
	\foreach \x in {1,2, ..., 9}{
		\foreach \y in {1, 2, ..., 4}{
			\draw[fill = black!10] (\x, #1, \y) -- (\x, #1, \y+1);
		}
	}
	%pes
	\foreach \x in {1,2, ..., 9}{
		\foreach \y in {1, 2, ..., 5}{
			\draw[fill = black!10] (\x-.2, #1, \y - .2) -- (\x-.2, #1, \y + .2) -- (\x+.2, #1, \y + .2) -- (\x+.2, #1, \y - .2) -- cycle;
		}
	}
}

\newcommand{\drawConnections}[2]{
	\foreach \x in {1,2, ..., 9}{
		\foreach \y in {1, 2, ..., 5}{
			\draw[fill = black!10] (\x, #1, \y) -- (\x, #2, \y);
		}
	}
}

\newcommand{\drawDataUp}[1]{
\draw[col2, fill = col2!50, opacity = .8] (0,#1, 8) -- (0,#1,12) --(10,#1,12) -- (10,#1,8) -- cycle;% opacity= 0.1
\draw[col2, -latex] (5, #1, 8) -- (5, #1, 6);
}
\newcommand{\drawDataUpConn}[2]{
	\foreach \x in {0, 10}{
		\foreach \y in {8, 12}{
			\draw[col2, densely dotted] (\x, #1, \y) -- (\x, #2, \y);
		}
	}
}

\newcommand{\drawDataLeft}[1]{
	\draw[col3, fill = col3!50, opacity = .8] (-2,#1, 0) -- (-2,#1,6) --(-6,#1,6) -- (-6,#1,0) -- cycle;% opacity= 0.1
	\draw[col3, -latex] (-2, #1, 3) -- (0, #1, 3);
}
\newcommand{\drawDataLeftConn}[2]{
	\foreach \x in {-2, -6}{
		\foreach \y in {0, 6}{
			\draw[col3, densely dotted] (\x, #1, \y) -- (\x, #2, \y);
		}
	}
}

\newcommand{\drawDataOut}[1]{
	\draw[red, fill = red!50] (0,#1, -2) -- (0,#1,-6) --(10,#1,-6) -- (10,#1,-2) -- cycle;% opacity= 0.1
	\draw[red, -latex] (5, #1, 0) -- (5, #1, -2);
}

\tikzmath{\l1 = 0;};
\tikzmath{\l2 = 1.3;};
\tikzmath{\l3 = 2*1.3;};

%DRAM
\draw[latex-latex, double] (-8.5,\l3, -1) -- (-7,\l3,-1);
\draw[latex-latex, double] (-8.5,\l3, -2) -- (-7,\l3,-2);
\draw[latex-latex, double] (-8.5,\l3, -3) -- (-7,\l3,-3);
\draw[fill = OliveGreen!95] (-8.5,\l3,-7) -- (-11.5,\l3,-7) -- (-11.5,\l3,3) -- (-8.5,\l3,3) -- cycle;
\foreach \x in {0.5, 3, ..., 9}{
	\fill[black] (-11+1.5,\l3,\x-7) -- (-12.5+1.5,\l3,\x-7) -- (-12.5+1.5,\l3,\x+1.5-7) -- (-11+1.5,\l3,\x+1.5-7) -- cycle;	
}
\foreach \x in {0.5, 5.5}{
	\draw[fill = Goldenrod] (-10+1.5,\l3,\x-7) -- (-10.5+1.5,\l3,\x-7) -- (-10.5+1.5,\l3,\x+4-7) -- (-10+1.5,\l3,\x+4-7) -- cycle;
}
\foreach \y in {0.5, 1, ..., 4.5}{
	\draw[] (-10+1.5,\l3,\y-7) -- (-10.5+1.5,\l3,\y-7);
}
\foreach \y in {5.5, 6, ..., 9.5}{
	\draw[] (-10+1.5,\l3,\y-7) -- (-10.5+1.5,\l3,\y-7);
}
\node[]() at(-11.5+1.5,\l3,4.5) {\scriptsize DRAM};
\draw[fill = black!5] (-7,\l3,-1) -- (11,\l3,-1) -- (11,\l3,13) -- (-7,\l3,13) -- cycle;
%hor. DRAM
%\draw[fill = OliveGreen!95] (-2,\l3,-4) -- (-2-10,\l3,-4) -- (-2-10,\l3,-4-3) -- (-2,\l3,-4-3) -- cycle;
%\foreach \x in {-2.5, -5, ..., -11}{
%	\fill[black] (\x,\l3, -6.5) -- (\x-1.5,\l3,-6.5) -- (\x-1.5,\l3,-6.5+1.5) -- (\x,\l3,-6.5+1.5) -- cycle;	
%}
%\foreach \x in {-2.5, -7.5}{
%	\draw[fill = Goldenrod] (\x,\l3,-4) -- (\x-4,\l3,-4) -- (\x-4,\l3,-4-.5) -- (\x,\l3,-4-.5) -- cycle;
%}
%\foreach \x in {-2.5,-2.75,..., -6.5}{
%	\draw[fill = Goldenrod] (\x,\l3,-4) -- (\x,\l3,-4-.5);
%}
%\foreach \x in {-7.5,-7.75,..., -11.5}{
%	\draw[fill = Goldenrod] (\x,\l3,-4) -- (\x,\l3,-4-.5);
%}
%\node[]() at(-10,\l3,-2) {\scriptsize DRAM};
%\foreach \x in {-2, -3, -4}{
%	\draw[latex-latex, double, thick] (\x,\l3,-3) -- (\x,\l3,-3-.5)
%}

%base die
\draw[fill = black!5] (-7,\l3,-7) -- (11,\l3,-7) -- (11,\l3,13) -- (-7,\l3,13) -- cycle;

\drawLayer{\l3}
\drawDataUp{\l3}
\drawDataLeft{\l3}
\drawDataOut{\l3}
\drawDataUpConn{\l2}{\l3}
\drawDataLeftConn{\l2}{\l3}
\drawConnections{\l2}{\l3}

\drawLayer{\l2}
\drawDataUp{\l2}
\drawDataLeft{\l2}
\drawDataUpConn{\l2}{\l1}
\drawDataLeftConn{\l2}{\l1}
\drawConnections{\l2}{\l1}

\drawLayer{\l1}
\drawDataUp{\l1}
\drawDataLeft{\l1}

\draw[blue, fill= blue!30] (8.5,\l3, -4.5) -- (8.5,\l3,-5.5) --(9.5,\l3,-5.5) -- (9.5,\l3,-4.5) -- cycle;
\foreach \x in {9}{
	\foreach \y in {1}{
		\draw[blue, -latex, rounded corners] (\x, \l1, \y) -- (\x, \l3, \y) -- (\x, \l3, \y-6);
	}
}

\node[](l1) at (5,\l1, 10) {\scriptsize Input A (SRAM)};
\node[align = center](l2) at (-4,\l1, 3) {\scriptsize Input B\phantom{33} \\ \hspace{10pt}\scriptsize (SRAM)};
\node[](l3) at (5,\l3, -4) {\scriptsize Output (SRAM)};
\node[align = left, anchor = south west](l4) at (12,\l1, 8) {\scriptsize 3D-systolic array};
\draw[-latex] (l4) -- (10,\l1,6);
\node[align = left, anchor = west](l5) at (13,\l1, 3) {\scriptsize MAC-unit};
\draw[-latex] (l5) -- (9.1,\l1,3.1);
\node[align = left, anchor = west](l6) at (12,\l2, 0) {\scriptsize tiers};
\draw[-latex] (l6) -- (10,\l1,0);
\draw[-latex] (l6) -- (10,\l2,0);
\draw[-latex] (l6) -- (10,\l3,0);

\end{tikzpicture}

%\end{document}
}
	\caption{ Schematic depicting the construction and data movement in a 3D systolic array.}
	\label{fig:3d}
\end{figure}
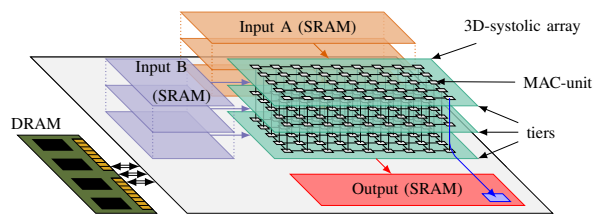

We extend the accelerator to the third spatial dimension to increase its parallelism since the runtime can be reduced by remapping.
Fig.~\ref{fig:3d} depicts the resulting 3D-accelerator, comprising layers of 2D systolic arrays stacked in 3D. The connections between the Multiply Accumulate Units (MACs) across the third dimensions (\textit{'tiers'}) enable the entire 3D structure to work as a single unit. 
If there are no connections between the tiers, or these connections are not used during computation, then the accelerator works as a scaled-out 2D system, implemented using 3D technology. 

Here,  we study the impact of 3D-DNN accelerator in terms of computational and thermal performance, power, scalability, and area. 
We choose a logic-stacked systolic-array architecture for the sake of simplicity.
We deliberately choose compute mappings that require communication across \textit{tiers} so that the architecture is not equivalent to a scaled-out 2D system.
%Our results will depict that optimal acceleration can be achieved by increasing the number of \textit{tiers} (see Fig.~\ref{fig:optimaltiers}). 

3D-Integrated Circuits (ICs) are fabricated either as a stacked 3D-IC, vertically interconnecting with Through-silicon-vias (TSVs), or a monolithic 3D-IC, vertically interconnecting with monolithic intertier vias (MIVs). Employing 3D-technology is challenging because of reduced yield vs. 2D, large area requirements of TSVs, and severe thermal limitations \cite{Perry.2019}. There are worthy benefits, e.g., less power consumption vs. 2D~\cite{Dong.2009}; Even fundamental limits of computation are tackled \cite{Markov.2014}. Therefore, 3D integration is being introduced by industry at time of writing this paper: For example, Intel ‘‘Lakefield’’, in which Foveros 3D technology is used to stack multicore processors and DRAM \cite{Intel.2019a}. However, there is still a lack of research on the advantages of 3D integration on DNN-accelerators.

To investigate 3D integration for DNN accelerators, we created a 3D-systolic array analytical model. Our results shows that a 3D-accelerator gets a 9.14$\times$ speedup compared with a 2D one with same number of compute units. 
We also design a 3D systolic array in RTL and perform post-synthesis area, power, and thermal analyses. 
We show that performance benefits for 3D are nullified by area costs in some configurations. The power analysis depicts that the array consumes similar amount of power as in 2D; while the thermal performance of 3D is worse than 2D, it is still feasible.

In this paper we claim the following contributions:
%\squishlist
\begin{itemize}[nosep]
\item{We extend an analytical performance model from 2D to 3D that allows to optimize parameters such as tier count, computational resources and array dimensions for a given workload.}
\item{We conduct a power, performance, and area (PPA) analysis from an RTL implementation for real workloads comparing 2D against 3D with TSVs and 3D with MIV.}
\item{We use the RTL model to conduct a thermal study. 2D is compared vs. 3D (TSV and MIV).}
\end{itemize}
%squishend

%The remainder of the paper is ordered as follows: In Sec.~\ref{sec:3d-arrays}, we introduce the architecture of our 3D array, a 3D-dataflow and the analytical performance model. In Sec.~\ref{sec:results}, we discuss the performance, power and area (PPA) and thermal properties of 3D for DNN-accelerators. We also discuss design implications on the tier count, required computational resources and favorable workload characteristics. In Sec.~\ref{sec:related}, we given an overview on related works. We conclude in Sec.~\ref{sec:conclusion}.
%\vspace{-8pt}
\section{Related Work}\label{sec:related}
%\vspace{-2pt}

Kung et al. \cite{Kung.2019} propose a 3D systolic array by folding a 2D one. Dataflows are essentially equivalent to 2D within each tier. This results in performance advantages from: A scale-out approach, reduced link lengths comparing TSVs to wires and concatenating the execution of NN layers. Unlike this work, \cite{Kung.2019} does not utilize 3D to leverage spacial parallelism in the temporal domain. MIVs are not analyzed.

Wang et al. \cite{Wang.2019} propose a systolic cube. The 3D topology is identical this work. The key differences are: the cube is implemented on a 2D IC and therefore is limited in scalability from wire length; the proposed dataflow targets 3D convolution. Our approach is more general such that any matrix multiplication is paralellized exploiting the extra spatial dimension enabled by the 3D substrate. The dataflow of \cite{Wang.2019} can be used in our 3D-IC, as well. Rahman et al. \cite{Rahman.2016} propose a logical 3D array implemented on a 2D-IC. MACs within a layer (loosely analog to tiers) are not connected as each MAC maps one neuron.

TETRIS \cite{Gao.2017} implements a NN accelerator in the logic layer of 3D stacked memory. The array is a 2D topology. The approach saves memory traffic and reduces power. 

Lakhani et al. \cite{Lakhani.1996} parallelize matrix multiplication in 3D. Although conceptually similar, \cite{Lakhani.1996} does not target the specific requirements of NN. Furthermore, the parallelization scheme along the third dimension relays on splitting up operations for floating point arithmetic among \textit{tiers}, but each \textit{tier} gets the whole set of inputs. Their proposed performance model is limited to matrices as big as the systolic array. The 3D array is not implemented; reliable area, power and thermal figures are not presented. 
Furthermore, a comparison of TSV and MIV-based implementations is not given.

\vspace{\sectionRemoveSpace}
\section{3D Systolic Array for NNs}\label{sec:3d-arrays}

\vspace{\subsectionRemoveSpaceUnderSection}
\subsection{Architecture}\label{sec:3d-arrays:arch}
%\vspace{-2pt}

The architecture of our 3D systolic array is shown in Fig.~\ref{fig:3d}. The MACs are connected to direct neighbors; horizontally via wires and vertically via TSVs/MIVs. The whole array calculates a matrix multiplication. Matrices can enter from top and left, shown in purple and orange. The matrix multiplication is parallelized in the third dimension by splitting it into partial sums. Thus, each layer is entered by two corresponding parts of the input matrices. The MAC units in the layer generate a partial sum. At the end, each pile of stacked MACs accumulates the data; then, the bottom layer returns the output matrix. (Other schemes would not require vertical communication, and our system was equivalent to a scaled-out accelerator.) This is shown in blue in Fig.~\ref{fig:3d} for one exemplary MAC pile. Only minor modifications to the MAC unit in comparison to a 2D array are necessary: One MUX, the accumulate control signal (partial summing across layers) and the vertical links are added. 

Please note that we connect each pair of adjacent MACs with a TSV/MIV array between layers. In general, this is a over-provision of vertical interconnects that induces an area overhead (especially for TSVs) and reduces the chip's yield. However, we chose this deliberately, as it is a worst-case approximation for 3D DNN-accelerators. Many methods that reduce the area overhead and increase the yield by limiting the number of vertical links exist in the literature
%most of them exploit the fact that vertical interconnects are very short and have a very small delay on comparison to 2D links. E.g., TSV arrays serialization can be applied to share TSV arrays
(e.g., \cite{Darve.2011}). We do not discuss these here, as it is outside of the scope of this paper and existing approaches can be applied.

\vspace{\subsectionRemoveSpace}
\subsection{Memory}
%\vspace{-2pt}
%To feed data into the 3D-array, it must be connected with memory: 
%%
%First, the data must be fetched from main memory into the array. In general, this works similar to a standard 2D-IC, ie. DRAM is connected with the array on one layer. Then, the data are distributed to the layers. 
The 3D accelerator talks to the external memory using a memory controller connected in one of the array's layers. 
The requested data coming from outside of the chip is distributed to the layers in a fashion similar to one in 
%This has already been analyzed by 
\cite{Kung.2019},
%so that we do not conduct a further study of this topic. 
%The authors of \cite{Kung.2019} compare a 2D-IC (scaled-out) with its 3D-IC counterpart
in which the small systolic arrays are distributed in 3D among dies \cite[Fig. 7]{Kung.2019}. The 3D array profits of a one to two orders of magnitude smaller wire delay for these vertical interconnects \cite[Fig. 8]{Kung.2019}. 

Data are stored intermediately before being fed into the array in scratchpad memory (cf. \cite[Fig. 1]{Samajdar.2019}). There are two options: Either, there is SRAM on one tier that interconnects to all tiers; or each tier has dedicated SRAM. Both are possible due to the small wire delay for 3D \cite{Kung.2019}. As the size and architecture of the scratchpad memory has a vast influence on the performance, this has already been optimized for 2D, e.g., \cite{Samajdar.2019}. Those findings can also be applied here. Hence, the architecture and the parameters of scratchpad memory are outside of the scope of this paper.

To summarize, 
%we deliberately do not consider connection to main memory and the layers' scratchpad memory, as both topics already have been taken into consideration by other works.
we do not claim any contribution in the memory system for a 3D systolic array and refer to the existing architectures proposed in the recent literature.
For main memory, \cite{Kung.2019} proved a performance advantage of 3D-ICs. For scratchpad memory, the findings of 2D-ICs can be applied, as each tier has dedicated memory. 

\vspace{\subsectionRemoveSpace}
\subsection{Mapping/Dataflow}
%\vspace{-2pt}

\begin{figure}
	\centering
	\scalebox{.55}{%\documentclass[border=2mm]{standalone}
%\usepackage[dvipsnames]{xcolor}
%\usepackage{pgfplots}
%\pgfplotsset{compat=1.8}
%\usepackage{amssymb}
%\usepackage{units}
%\usetikzlibrary{decorations.pathreplacing, decorations, positioning, calc, patterns, arrows, math, shadings}
%\input{defs.tex}

%\begin{document}
\begin{tikzpicture}[scale = .3, yscale = -1]

%MACs
\foreach \x in {0,3,6, 9}{
	\foreach \y in {0,3,6}{
		\draw[fill = black!10] (\x-1,  \y- 1) -- (\x-1,  \y+ 1) -- (\x+1,  \y+ 1) -- (\x+1, \y- 1) -- cycle;
		\node[anchor =center] (mac) at (\x,\y) {\normalsize $\times$+};
		\draw[-latex, NavyBlue] (\x+.6,  \y-1) to[out=-90,in=0, distance = 1.2cm] (\x+1,  \y-.6);
	}	
}
\node[anchor = west, align = left,  NavyBlue](n1) at (9+1,0+.2) {\normalsize inplace\\[-2pt]\normalsize reduction};
%Links
\foreach \x in {0,3,6}{
	\foreach \y in {0,3,6}{
		\draw[densely dotted, -latex] (\x+1, \y) -- (\x + 2, \y);
	}	
}
\foreach \x in {0,3,6, 9}{
	\foreach \y in {0,3}{
		\draw[densely dotted, -latex] (\x, \y+1) -- (\x, \y+2);
	}	
}

%data top:
\draw[col2, fill = col2!10] (0-1, -0/3-2) rectangle ++(2, -6);
\draw[col2, fill = col2!30] (3-1, -3/3-2) rectangle ++(2, -6);
\draw[col2, fill = col2!50] (6-1, -6/3-2) rectangle ++(2, -6);
\draw[col2, fill = col2!70] (9-1, -9/3-2) rectangle ++(2, -6);
\foreach \x in {0,3,6, 9}{
%	\draw[col2, fill = col2!10] (\x-1, -\x/3-2) rectangle ++(2, -6);
	\draw [-latex] (\x, -\x/3-2) -- (\x, -1); 	
}

%data left:
\draw[col3, fill = col3!10] (-0/3-2, 0-1) rectangle ++(-6, 2);
\draw[col3, fill = col3!40] (-3/3-2, 3-1) rectangle ++(-6, 2);
\draw[col3, fill = col3!60] (-6/3-2, 6-1) rectangle ++(-6, 2);
\foreach \y in {0,3,6}{
	\draw [-latex] (-\y/3-2, \y) -- (-1,\y); 	
}

%%matrices:
\coordinate (mA) at (-11,-3);
\draw[col2, fill = col2!10] ($(mA) + (0,0)$) rectangle ++(2, -6);
\draw[col2, fill = col2!30] ($(mA) + (2,0)$)  rectangle ++(2, -6);
\draw[col2, fill = col2!50] ($(mA) + (4,0)$)  rectangle ++(2, -6);
\draw[col2, fill = col2!70] ($(mA) + (6,0)$)  rectangle ++(2, -6);
\draw[] (mA) rectangle ++(8,-6);
\node[] () at ($(mA) + (4,-3)$) {\normalsize$A^{(M\times K)}$};
\draw[latex-latex] ($(mA) + (0,-6.5)$) -- ++ (8,0) node [midway, above] {\normalsize$M$};
\draw[latex-latex] ($(mA) + (-.5,0)$) -- ++ (0, -6) node [midway, left] {\normalsize$K$};

\coordinate (mB) at (-20,-3);

\draw[col3, fill = col3!60] ($(mB) + (0,0)$) rectangle ++(6, -2);
\draw[col3, fill = col3!40] ($(mB) + (0,-2)$) rectangle ++(6, -2);
\draw[col3, fill = col3!10] ($(mB) + (0,-4)$) rectangle ++(6, -2);
\draw[] (mB) rectangle ++(6,-6);
\node[] () at ($(mB) + (3,-3)$) {\normalsize$B^{(K\times N)}$};
\draw[latex-latex] ($(mB) + (0,-6.5)$) -- ++ (6,0) node [midway, above] {\normalsize$K$};
\draw[latex-latex] ($(mB) + (-.5,0)$) -- ++ (0, -6) node [midway, left] {\normalsize$N$};

\end{tikzpicture}

%\end{document}
}
	%\vspace{-6pt}
	\caption{Figure depicts the output-stationary dataflow for a conventional 2D systolic array.}
	\label{fig:dataflow2Dintro}
	\vspace{8pt}
	\centering
	\scalebox{1}{%\documentclass[border=2mm]{standalone}
%\usepackage[dvipsnames]{xcolor}
%\usepackage{pgfplots}
%\pgfplotsset{compat=1.8}
%\usepackage{amssymb}
%\usepackage{units}
%\usetikzlibrary{decorations.pathreplacing, decorations, positioning, calc, patterns, arrows, math, shadings}
%\input{defs.tex}
%
%\begin{document}
\begin{tikzpicture}[scale = .3, yscale = -1]

%draw MACs:
\newcommand{\drawMAC}[1]{
	\draw[fill = black!10] (-1, #1,  - 1) -- (-1, #1,  + 1) -- (+1, #1,  + 1) -- (+1, #1,  - 1) -- cycle;
	\node[anchor =center] (mac) at (0,#1, 0) {\tiny $\times$+};
	\draw[-latex, NavyBlue] (.6, #1,  1) to[out=-90,in=0, distance = 1.2cm] (1, #1,  .6);
	\node[anchor = west, NavyBlue](n1) at (1.5,#1,1) {\tiny inplace reduction};
}
%#1 layer, #2 skew
\newcommand{\dataflowTop}[2]{
	\draw[col2, fill = col2!10] (-1, #1, 3+#2) -- (1, #1, 3+#2) -- (1, #1, 10+#2) -- (-1, #1, 10+#2) --cycle;
	\draw[col2, fill = col2!50] (-1, #1, 3+#2) -- (1, #1, 3+#2) -- (1, #1,  5+#2) -- (-1, #1,  5+#2) --cycle;
	\draw[-latex, col2] (0, #1, 3+#2) -- (0,#1, 1);
}

%#1 layer, #2 skew
\newcommand{\dataflowLeft}[2]{
	\draw[col3, fill = col3!10] (-3-#2, #1, -1) -- (-3-#2, #1, 1) -- (-10-#2, #1, 1) -- (-10-#2, #1, -1) --cycle;
	\draw[col3, fill = col3!50] (-3-#2, #1, -1) -- (-3-#2, #1, 1) -- ( -5-#2, #1, 1) -- ( -5-#2, #1, -1) --cycle;
	\draw[-latex, col3] (-3-#2, #1, 0) -- (-1,#1, 0);
}

\tikzmath{\l4 = 2.5*1.8;};
\tikzmath{\l1 = 0;};
\tikzmath{\l2 = 1.8;};
\tikzmath{\l3 = 2*1.8;};

%MACs and their vertical connections
\drawMAC{\l3}
\draw[-latex, Plum] (0,\l2,0) -- (0,\l3,0);
\node[anchor = west, Plum]() at (0,.65*\l2+.35*\l3, 0)  {\tiny reduce time across tiers};
\dataflowTop{\l3}{4}
\dataflowLeft{\l3}{6}

\drawMAC{\l2}
\draw[-latex, Plum] (0,\l1,0) -- (0,\l2,0);
\node[anchor = west, Plum]() at (0,.65*\l1+.35*\l2, 0)  {\tiny reduce time across tiers};
\dataflowTop{\l2}{2}
\dataflowLeft{\l2}{4}

\drawMAC{\l1}
\dataflowTop{\l1}{0}
\dataflowLeft{\l1}{2}

\draw[densely dotted, OrangeRed](-1, \l1, 3) -- (-1, \l4, 3);
\draw[latex-, OrangeRed](-1, \l4, 3) -- (-1, \l4, 5) node [midway, above] {\tiny d};
\draw[densely dotted, OrangeRed](-1, \l1, 5) -- (-1, \l4, 5);
\draw[latex-, OrangeRed](-1, \l4, 5) -- (-1, \l4, 7) node [midway, above] {\tiny d};
\draw[densely dotted, OrangeRed](-1, \l2, 7) -- (-1, \l4, 7);

\draw[densely dotted, OrangeRed](-5, \l1, -1) -- (-5, \l4, -1);
\draw[-latex, OrangeRed](-7, \l4, -1) -- (-5, \l4, -1) node [midway, above] {\tiny d};
\draw[densely dotted, OrangeRed](-7, \l1, -1) -- (-7, \l4, -1);
\draw[-latex, OrangeRed](-9, \l4, -1) -- (-7, \l4, -1) node [midway, above] {\tiny d};
\draw[densely dotted, OrangeRed](-9, \l2, -1) -- (-9, \l4, -1);

\node[anchor = east, OrangeRed] () at (-11, \l4, -1) {\tiny d = 1 cycle};

\draw[-latex, BrickRed](0,\l3, -1) -- (0,\l3, -2.5) node [anchor = north, align = center] {\tiny final output};

\end{tikzpicture}

%\end{document}
}
	%\vspace{-11pt}
	\caption{Figure depicts the distributed output-stationary dataflow for a single pile of MACs (ranging over three tiers).}
	\label{fig:dataflowMAC}
	\vspace{8pt}
	\centering
	\scalebox{1}{%\documentclass[border=2mm]{standalone}
%\usepackage[dvipsnames]{xcolor}
%\usepackage{pgfplots}
%\pgfplotsset{compat=1.8}
%\usepackage{amssymb}
%\usepackage{units}
%\usetikzlibrary{decorations.pathreplacing, decorations, positioning, calc, patterns, arrows, math, shadings}
%\input{defs.tex}
%
%\begin{document}
\begin{tikzpicture}[scale = .3, yscale = -1]

%draw MACs:
\newcommand{\drawMAC}[1]{
	\draw[black, fill =black!3, opacity = .8] (-16.5, #1,  -5) -- (-16.5, #1,  17) -- (5, #1,  17) -- (5, #1,   -5) -- cycle;
	\foreach \x in {-3, 3}{
		\foreach \y in {-3,3}{
			\draw[fill = black!10] (\x-1, #1,  \y- 1) -- (\x-1, #1,  \y+ 1) -- (\x+1, #1,  \y+ 1) -- (\x+1, #1,  \y- 1) -- cycle;
			\node[anchor =center] (mac) at (\x,#1,\y) {\tiny $\times$+};
		}
	}
	\foreach \x in {-3, 3}{
		\foreach \y in {-3}{
			\draw[] (\x, #1,  \y+1) -- (\x, #1,  \y+5);
		}
	}
	\foreach \x in {-3}{
		\foreach \y in {-3, 3}{
			\draw[] (\x+1, #1,  \y) -- (\x+5, #1,  \y);
		}
	}
}
%#1 layer, #2 skew
\newcommand{\dataflowTop}[3]{
	\foreach \x in {-3}{
		\draw[col2, fill = col2!10, opacity = #3] (\x-1, #1, 3+#2) -- (\x+1, #1, 3+#2) -- (\x+1, #1, 9+#2) -- (\x-1, #1, 9+#2) --cycle;
		\draw[col2, fill = col2!50, opacity = #3] (\x-1, #1, 3+#2) -- (\x+1, #1, 3+#2) -- (\x+1, #1,  5+#2) -- (\x-1, #1,  5+#2) --cycle;
		\draw[-latex, col2, opacity = #3, thick] (\x, #1, 3+#2) -- (\x,#1, 4);
	}
	\foreach \x in {3}{
		\draw[col2, fill = col2!10, opacity = #3] (\x-1, #1, 3+#2+2) -- (\x+1, #1, 3+#2+2) -- (\x+1, #1, 9+#2+2) -- (\x-1, #1, 9+#2+2) --cycle;
		\draw[col2, fill = col2!50, opacity = #3] (\x-1, #1, 3+#2+2) -- (\x+1, #1, 3+#2+2) -- (\x+1, #1,  5+#2+2) -- (\x-1, #1,  5+#2+2) --cycle;
		\draw[-latex, col2, opacity = #3, thick] (\x, #1, 3+#2+2) -- (\x,#1, 4);
	}
%	\draw[red, densely dotted] () -- ();
}

%#1 layer, #2 skew
\newcommand{\dataflowLeftFront}[2]{
	\foreach \y in {-3}{
		\draw[col3, fill = col3!10] (-3-#2, #1, \y-1) -- (-3-#2, #1, \y+1) -- (-9-#2, #1, \y+1) -- (-9-#2, #1, \y-1) --cycle;
		\draw[col3, fill = col3!50] (-3-#2, #1, \y-1) -- (-3-#2, #1, \y+1) -- ( -5-#2, #1, \y+1) -- ( -5-#2, #1, \y-1) --cycle;
		\draw[-latex, col3, thick] (-3-#2, #1, \y) -- (-4,#1, \y);
	}
}
\newcommand{\dataflowLeftBack}[3]{
	\foreach \y in {3}{
		\draw[col3, fill = col3!10, opacity = #3] (-3-#2+2, #1, \y-1) -- (-3-#2+2, #1, \y+1) -- (-9-#2+2, #1, \y+1) -- (-9-#2+2, #1, \y-1) --cycle;
		\draw[col3, fill = col3!50, opacity = #3] (-3-#2+2, #1, \y-1) -- (-3-#2+2, #1, \y+1) -- ( -5-#2+2, #1, \y+1) -- ( -5-#2+2, #1, \y-1) --cycle;
		\draw[-latex, col3, opacity = #3, thick] (-3-#2+2, #1, \y) -- (-4,#1, \y);
	}
}
\newcommand{\dataflowLeft}[2]{
	\dataflowLeftFront{#1}{#2}
	\dataflowLeftBack{#1}{#2}{1}
}

\tikzmath{\l4 = 2.5*1.8;};
\tikzmath{\l1 = 0;};
\tikzmath{\l2 = 1.8;};
\tikzmath{\l3 = 2*1.8;};

\drawMAC{\l3}
\node[] () at (.2,\l3, 0) {\tiny bottom tier};
\dataflowTop{\l3}{6}{.9}
\dataflowLeftFront{\l3}{7}
\dataflowLeftBack{\l3}{7}{.3}
\foreach \x in {-3, 3} {
	\foreach \y in {-3, 3}{
		\draw[] (\x, \l3, \y) -- (\x, \l1, \y);
	}
}

\drawMAC{\l1}
\node[] () at (.2,\l1, 0) {\tiny upper tier};
\dataflowTop{\l1}{4}{1}
\dataflowLeft{\l1}{5}

%red desciptions
%top
\foreach \x in {7, 9}{
	\draw[OrangeRed, densely dotted] (4, \l1, \x) -- (-5.5, \l1, \x);
}
\draw[latex-, OrangeRed] (-5.5, \l1, 7) -- (-5.5, \l1, 9) node [midway, right ] {\tiny d};

%left
\foreach \x in {-6, -8}{
	\draw[OrangeRed, densely dotted] (\x, \l1, -4) -- (\x, \l1, 5.5);
}
\draw[latex-, OrangeRed] (-6, \l1, 5.5) -- (-8, \l1, 5.5) node [midway,above, yshift = -3] {\tiny d};

%between two layers
\foreach \x in {-8, -10}{
	\draw[OrangeRed, densely dotted] (\x, \l1, -4) -- (\x, \l4, -4);
}
\draw[latex-, OrangeRed] (-8, \l4, -4) -- (-10, \l4, -4) node [midway,below] {\tiny d};

%SRAM desc
\node[] () at (-4,\l1,15) {\tiny Filter fed from SRAM};
\node[rotate = -45] () at (-14,\l1,6) {\tiny IFMAP fed from SRAM};

\end{tikzpicture}

%\end{document}
}
	%\vspace{-22pt}
	\caption{Figure depicting the distributed output-stationary dataflow for multiple MACs within a tier and 2 tiers.}
	\label{fig:dataflowMACMultiple}
\end{figure}
%
%\begin{figure}
%	\centering
%	\input{fig/3darrayTiming}
%	\caption{\moritz{NAME} dataflow to 3D array \moritz{Other figures required... ideas?}}
%	\label{fig:3dOS}
%\end{figure}

% Anand: Reworded the following
\begin{comment}
    Systolic arrays can map computations in three major strategies: output stationary (OS), weight stationary (WS) and input stationary (IS). 
    Chen et al. \cite{eyeriss} describe these strategies or dataflows. 
    The OS dataflow is the strategy where each MAC unit is responsible for computing a given output feature map (OFMAP) pixel. All partial sums are generated and reduced in place within the given MAC, while the necessary operands are sent to the MAC over the links of the systolic array.
    In WS dataflow elements of the filter matrix are first distributed over the  array such that one element is stored in every MAC. Elements corresponding to a single filter are mapped over a single column. The input feature map (IFMAP) elements are streamed to the array and communicated over the links in a systolic fashion. 
    Each MAC calculates one partial sum per cycle, which are then passed across rows along  columns for reduction. 
    In IS dataflow the mapping and reduction strategy is similar to WS, except the elements of IFMAP matrix are stored in the MAC units and elements of the filter matrix are streamed to the array.
\end{comment}
% 
The mapping of operands plays an important role in determining the performance of a workloads on an accelerator. 
Chen et al. \cite{eyeriss} describe the various mapping strategies in DNN accelerators and introduce a naming convention used here.
\cite{Samajdar.2019} shows that out of the various strategies, three ones lend themselves for efficient mapping of computation on systolic arrays. These are output stationary (OS), weight stationary (WS) and input stationary (IS). 
In the following paragraphs we briefly describe these in the context of mapping a General Matrix Matrix Multiplication (GEMM) of two operand matrices $A^{(M\times K)}$ and $B^{(K\times N)}$, and discuss the implications when mapped onto a 3D systolic architecture.

\textbf{WS and IS dataflows.} 
In the WS dataflow the elements of the matrix $B$ are first stored to the local memory of the MAC units, such that each column of the array gets the element corresponding the a specific col of the matrix B (given sufficient MACs). 
Once the elements are stored, one element of a row of the transposed matrix of $A$ is fed in from the left side of the array in each cycle. 
Each MAC then multiplies the incoming element with its stored operand, and forwards incoming value to the MAC unit on its right. 
The generated product is then added with the incoming sum from the top edge, and the partial sum is sent to the element on the bottom row.
Thus, the elements corresponding to each column of the output are generated, within one column of the array, by reducing across the rows.
\textit{To summarize this strategy, the dimension $N$ is mapped spatially along the columns, the dimension $K$ is also mapped spatially along the rows. However, the dimension $M$ is temporally mapped.}

The IS dataflow is similar to WS, but the mapping of matrices $A$ and $B$ are interchanged.
The elements of each row in matrix $A$ are first stored into the local memory of MACs along array columns. 
The columns of matrix $B$ are then fed in from the left, one column per cycle, and the multiplication and reduction takes place similar to the case in WS. 
\textit{Thus, the dimension $M$ is mapped spatially along the array columns. 
The dimension $K$ is also mapped spatially along the array rows, while the dimension $N$ is mapped temporally.}

\textbf{OS dataflow.} 
The elements of matrix $B$ are streamed from the top edge of the array, such the each column of this matrix send to a particular array column; while the elements of matrix $A$ are streamed from the left edge of the array such that each array row receives elements from the corresponding row of the operand.
The partial sums are generated in each MAC and are reduced locally, producing the output matrix.
Fig.~\ref{fig:dataflow2Dintro} depicts the schematics of this mapping.
\textit{To summarize, the dimensions $M$ and $N$ are mapped along the array's spatial dimensions. The dimension $K$ is mapped along the temporal dimension.}

\textbf{Exploiting the third dimension.} 
The analysis above shows that the temporal dimension contributes to increase in runtime and therefore impedes performance (given an optimal 2D mapping). 
Adding a third spatial dimension can alleviate this bottleneck. 
In the context of WS and IS dataflow, this translates to mapping the $M$ and $N$ dimension in the `\textit{new}' spatial dimension. 
E.g., if we have a 3D stacked architecture, of 2 planar arrays; in WS dataflow, half of the rows in matrix $A$ would be used in the `\textit{top}' tier array, while the other half in the `\textit{bottom}' tier array. 
In case of IS dataflow, the mapping would be similar, but the roles of matrix $A$ and $B$ would be interchanged.
Please note that there is no communication between the arrays on the different tiers.
This is identical to a distributed array, and such acceleration lends itself into the well studied model parallelism approach \cite{Samajdar.2019}.

The OS dataflow, however, is an interesting one for 3D. 
Tthe dimension $K$ will be mapped to the third spatial dimension, therefore leading to reductions to be performed across the tiers. 
Fig.~\ref{fig:dataflowMAC} depicts the computation flow equivalent to a single MAC on a 2D array, working with OS strategy on the proposed 3D setting.
We refer to this new strategy as ``\textit{distributed output stationary (dOS)}".
In Fig.~\ref{fig:dataflowMACMultiple} we show a schematic of a 2-tiered 3D array employing \textit{dOS} dataflow. 
The mapping and reduction across various tiers lead to interesting architectural trade-offs to improve performance. 
The combination of inplace and cross-tier reduction means that na\"{i}vely increasing the number of tiers will lead to increased reduction time and will hamper performance (see Sec. \ref{subsec:comp-perf}).
\textit{In the rest of this paper, we focus on the dOS dataflow and study the performance and implementation aspects in a 3D systolic array setting. 
We do not dive into details for the WS and IS dataflows as the existing literature on model parallelism provides detailed analysis for these cases \cite{eyeriss}.
}

\subsection{Analytical performance model}

In \cite{Samajdar.2019}, an analytical performance model has been proposed: 
For a 2D systolic array with $R$ rows and $C$ columns (i.e. $\mathcal{N} = RC$ MACs) and workload matrices of dimensions $M$, $N$ and $K$, \cite[Eq. (4)]{Samajdar.2019} gives the calculation time:
%{
%\setlength{\abovedisplayskip}{0pt}
%\setlength{\belowdisplayskip}{0pt}
%\setlength{\abovedisplayshortskip}{0pt}
%\setlength{\belowdisplayshortskip}{0pt}
\begin{align}
\tau_{2D} = (2 R + C + T -2) \lceil \nicefrac{M}{R}\rceil \lceil \nicefrac{N}{C} \rceil
\label{eq:2d}
\end{align}
%}
Given large matrix sizes such that the given 2D array could not map the entire computation at once, serialization is required.
%
%Fig.~\ref{fig:scaleup} depicts the method employed to serialize. 
%\todo{@Anand: pls double check the fig}
The dimension $M$ will be mapped across the rows, if in case the number of rows is insufficient. The entire mapping requires $\lceil \nicefrac{M}{R}\rceil$ steps.
Similarly dimension $N$ is mapped across the columns, leading to $\lceil \nicefrac{N}{C} \rceil$ steps to complete the mapping.
The total number of serial steps required is given by $\lceil \nicefrac{M}{R}\rceil \lceil \nicefrac{N}{C} \rceil$. 

%\begin{figure}
%\begin{minipage}[b]{.49\linewidth}
%    \centering
	%\scalebox{.7}{\input{fig/2DScaleUp.tex}}
    %\vspace{-12pt}
    %\caption{{Figure depicting serialization in case the entire computation cannot be mapped at one step. Adopted figure from \cite{Samajdar.2019}.}}
	%\label{fig:scaleup}
%\end{minipage}\hfill
%\begin{minipage}[b]{.49\linewidth}
	%\centering
	%\scalebox{.9}{\input{fig/2Ddataflow.tex}}
	%\vspace{-12pt}
	%\caption{{Figure depicting the movement of operands and results within a single mapping step for output-stationary dataflow. Adopted figure from \cite{Samajdar.2019}.}}
%	\label{fig:dataflow2D}
%\end{minipage}
%\end{figure}

%Fig.~\ref{fig:dataflow2D} depicts the time taken at each serialization step for an OS data flow execution. 
In each step of the serialization it takes ($R + C - 2$) cycles to fill the entire array, since the elements of IFMAP and Filter matrices are fed simultaneously.
In OS dataflow the computation for each OFMAP pixel is done in-place within a MAC unit, thus it requires $K$ cycles to generate one OFMAP pixel. 
Since multiple MAC units are running in parallel, the latency of computation can be hidden for all but one MAC unit, which gets the data at the end. 
This MAC takes another $K$ cycles after the array is filled. 
Once all the computation is finished, it takes another $R$ cycles to remove all the generated outputs from the array.
The term, ($2R + C + T - 2$) therefore indicates the runtime for a single serial step or \textit{fold}. The authors also propose an optimization method to find the optimal array sizes for a given workload that minimizes this runtime. 

The given formula naturally extends to a third dimension. Using the OS dataflow for 3D, the work among tiers is split up in $K$-dimension, i.e. along $K$ in the workload. The workload is not split up along $M$ and $N$. Thus, each of the $\ell$ tiers works on the partial sums with an input workload matrix dimension of $M$, $N$ and $T/\ell$. At the end, the partial sums are accumulated; this requires $\ell - 1$ additions. This yields the following formula for the runtime of a 3D systolic array with $R'$ rows per tier and $C'$ columns per tier:
%{
%\setlength{\abovedisplayskip}{0pt}
%\setlength{\belowdisplayskip}{0pt}
%\setlength{\abovedisplayshortskip}{0pt}
%\setlength{\belowdisplayshortskip}{0pt}
\begin{align}
\tau_{3D} = (2R'+C'+(\nicefrac{K}{\ell} + \ell-1) -2) \left\lceil \nicefrac{M}{R'}\right\rceil \left\lceil \nicefrac{N}{C'}\right\rceil \label{eq:model}
\end{align}
%}
Please note that the constraint for the MAC count changed as the 3D array has $\mathcal{N} = \ell R'C'$ MACs. Thus, the method from \cite{Samajdar.2019} can be applied to optimize the array dimensions for all tiers for the workload by changing the objective function to Eq.~\ref{eq:model} and using $\mathcal{N}/\ell$ MACs and a workload size of $M$, $N$ and $K/\ell$. To generate the distributed OS dataflow, each tier has the same array dimensions.

%\moritz{As we want to add some more insights: We are also finding optimal array dimensions, in which each tier has the same size. Maybe we highlight this optimization, as well? Just as in our paper on scaling. It might be something, the reader does not see in effort put into the work.} \todo{Essentially done, just double check}
\vspace{\sectionRemoveSpace}
\section{Results and Design Implications}\label{sec:results}

%To identify design implications for 3D-integration applied to DNN-accelerators, we conducted an in-depth analysis of the design space for computational and thermal performance, area and power.

\vspace{\subsectionRemoveSpaceUnderSection}
We synthesized our RTL implementation for \unit[15]{nm} nangate node (FreePDK15) \cite{Martins.2015} using Synopsys\textsuperscript{\textregistered} Design Compiler; power analysis was done with post-synthesis with Synopsys\textsuperscript{\textregistered} PrimeTime PX. The thermal analysis was done with HotSpot 6.0 \cite{Stan.2015}. For performance analysis, we sweep workload parameters and take the range of $M$, $N$ and $K$ from typical DNNs. Exemplary workload parameters are shown in Table~\ref{table:workloads}.
 
\begin{table}[t] 
\centering 
\setlength{\abovecaptionskip}{3pt}
\setlength{\belowcaptionskip}{0pt} 
\caption{Matrix dimensions for exemplary layers from current DNN workloads mapped to $M$, $N$ and $K$.}
\label{table:workloads} 
\scriptsize 
\begin{tabular}{p{2.5cm}|p{1cm}|p{.8cm}|p{.8cm}|p{.8cm}}
\toprule
 \bf Name & \bf Layer  &  \bf $M$ & \bf $K$ & \bf $N$ \\  
\midrule
\multirow{2}{2.5cm}{Resnet50 \cite{He.2015}}  &RN0 & 64 & 12100 & 147\\
 &RN1 & 512 & 784 & 128\\
\multirow{2}{2.5cm}{Google's neural mashine translation \cite{Wu.2016}} & GNMT0 & 128 & 4096 & 2048 \\
&GNMT1 & 320 & 4096 & 3072 \\
%&GNMT2 & 1632 & 1024	& 36548 \\
%&GNMT3 & 2048 & 32 & 4096 \\
\multirow{2}{2.5cm}{DeepBench \cite{Deepbench}}  & DB0	  & 1024 & 50000 & 16 \\ 
 & DB1	  & 35	 & 2560	& 4096 \\
\multirow{2}{2.5cm}{Transformer \cite{vaswani2017attention}}  &TF0	& 31999	& 84 & 1024 \\
&TF1	& 84 & 4096 & 1024 \\
%\multirow{2}{2.5cm}{Neural Collaborative Filtering \cite{He.2017}}  &NCF0 & 2048 & 128 & 1 \\
%&NCF1 & 256 & 2048 & 256 \\
\bottomrule
\end{tabular}
%\vspace{-12pt}
\end{table} 

\vspace{\subsectionRemoveSpace}
\subsection{Performance}
\label{subsec:comp-perf}

\begin{figure}
	\centering
	\scalebox{.9}{
	%\documentclass[border=2mm]{standalone}
%\usepackage[dvipsnames]{xcolor}
%\usepackage{pgfplots}
%\pgfplotsset{compat=1.8}
%\usepackage{amssymb}
%\usepackage{units}
%\usetikzlibrary{decorations.pathreplacing, decorations, positioning, calc, patterns, arrows, math, shadings, matrix}
%\input{defs.tex}

%\begin{document}
	\begin{tikzpicture}
		\begin{axis}[
		ymode=log,
		axis line style={lightgray},
		grid,
		ymax = 9.5,
		ymin = 0.5,
		xmin = -6,
		xmax = 12,
		title style = {font =\small, yshift =-5},
		label style={font=\scriptsize},
		ylabel style = {yshift = -15},
		xlabel style = {yshift = 5},
		ytick={.5, 1, 2, 4 , 8},
		yticklabels = {0.5, 0, 1, 2, 4},
		x tick label style={font=\scriptsize, yshift = 0},
		y tick label style = {font=\scriptsize},
		legend style = {font = \tiny},
   		legend pos = north west,
   		legend image post style={scale=0.4},
   		legend style={row sep=-4pt},
   		legend columns=3,
   		legend cell align={left},
	%	restrict y to domain*=0:1.5,
%			visualization depends on=rawy\as\rawy,
%			 after end axis/.code={ % Draw line indicating break
%				\draw [ultra thick, white, decoration={snake, amplitude=1pt}, decorate] (rel axis cs:0,1.05) -- (rel axis cs:1,1.05);
%			},
		xlabel = {tier count},
		ylabel = {speedup normalized to 2D},
		xtick= data,
		%symbolic x coords={1,2,3,4,5},
		height =4.4cm,
		width = 1.1\linewidth, 
		%		ymin= 0.9
	%	bar width=0.11cm,
		%every axis plot/.append style={fill,draw=none,no markers},
		ytick style={draw=none},
		xtick style={draw=none},clip=false,	]
		
		\coordinate (rect1) at (axis cs:-5.99,0.501);
		\coordinate (rect2) at (axis cs:11.99, 0.99);
		\fill[red!10] (rect1) rectangle(rect2);
		
		\addplot[myMarkerPlot, col1, mark = *] table[x=ell,y=mac4096k12100, col sep=comma]{data/xIsEll/plot1resultsSpeedup.csv};\label{plot:line1}
		\addplot[myMarkerPlot, col1, mark = triangle*] table[x=ell,y=mac4096k784, col sep=comma] {data/xIsEll/plot1resultsSpeedup.csv};\label{plot:line2}
		\addplot[myMarkerPlot, col1, mark = square*] table[x=ell,y=mac4096k255, col sep=comma] {data/xIsEll/plot1resultsSpeedup.csv};\label{plot:line3}

\addplot[myMarkerPlot, col2, mark = *] table[x=ell,y=mac32768k12100, col sep=comma] {data/xIsEll/plot1resultsSpeedup.csv};\label{plot:line21}
\addplot[myMarkerPlot, col2, mark = triangle*] table[x=ell,y=mac32768k784, col sep=comma] {data/xIsEll/plot1resultsSpeedup.csv};\label{plot:line22}
\addplot[myMarkerPlot, col2, mark = square*] table[x=ell,y=mac32768k255, col sep=comma] {data/xIsEll/plot1resultsSpeedup.csv};\label{plot:line23}

\addplot[myMarkerPlot, col3, mark = *] table[x=ell,y=mac262144k12100, col sep=comma] {data/xIsEll/plot1resultsSpeedup.csv};\label{plot:line31}
\addplot[myMarkerPlot, col3, mark = triangle*] table[x=ell,y=mac262144k784, col sep=comma] {data/xIsEll/plot1resultsSpeedup.csv};\label{plot:line32}
\addplot[myMarkerPlot, col3, mark = square*] table[x=ell,y=mac262144k255, col sep=comma] {data/xIsEll/plot1resultsSpeedup.csv};\label{plot:line33}

		\draw [thick, OrangeRed] ({rel axis cs:0,0}|-{axis cs:2,1}) -- ({rel axis cs:1,0}|-{axis cs:2,1}) node [pos=0.15, above, ForestGreen] {\sffamily \scriptsize $\Uparrow$ better} node [align = left, pos=0.15, below] {\sffamily \scriptsize $\Downarrow$ worse};
%		\legend{{MAC = 4096, K = 12100}, {MAC = 4096, K = 784}, {MAC = 4096, K = 255},
%		{MAC = 32768, K = 12100}, {MAC = 32768, K = 784}, {MAC = 32768, K = 255},
%		{MAC = 262144, K = 12100}, {MAC = 262144, K = 784}, {MAC = 262144, K = 255}};
	 \coordinate (legend) at (axis description cs:.0,.4);
		\end{axis}	
%	\matrix [
%	fill = white, 
%	draw,
%	every node/.append style={%
%		inner xsep=0pt,
%		inner ysep=0pt, 
%		draw=lightgray, % just to show node borders
%		text height=10,%\heightof{0},
%		text width=24,
%		align=center
%	},
%	matrix of nodes,
%	anchor=north west,
%	] at (legend) {
%		\tiny MAC\textbackslash K & \tiny \phantom{\textbackslash}12100 & \tiny \phantom{\textbackslash}784 & \tiny \phantom{\textbackslash}255\\
%		\tiny 4096 & \scalebox{1.5}{\ref{plot:line1}} & \scalebox{1.5}{\ref{plot:line2}} & \scalebox{1.5}{\ref{plot:line3}}   \\
%		\tiny 32768 & \scalebox{1.5}{\ref{plot:line21}} & \scalebox{1.5}{\ref{plot:line22}} & \scalebox{1.5}{\ref{plot:line23}}   \\
%		\tiny 262144 & \scalebox{1.5}{\ref{plot:line31}} & \scalebox{1.5}{\ref{plot:line32}} & \scalebox{1.5}{\ref{plot:line33}}   \\
%	};

	\node [anchor = south west ]at (legend) {\scalebox{.8}{
\begin{tabular}{|c|c|c|c|}

\hline 	
\cellcolor{white}& \multicolumn{3}{c|}{\cellcolor{white}\scriptsize\bfseries K}\\
\scriptsize\cellcolor{white}  \bfseries MACs & \scriptsize\cellcolor{white} 255 & \scriptsize \cellcolor{white}784 & \scriptsize \cellcolor{white}12100\\
\hline 
\scriptsize \cellcolor{white}$2^{12}$ & \cellcolor{white}\scalebox{1.5}{\ref{plot:line3}} & \cellcolor{white}\scalebox{1.5}{\ref{plot:line2}} &  \cellcolor{white}\scalebox{1.5}{\ref{plot:line1}}   \\
\hline 
\scriptsize \cellcolor{white}$2^{15}$ & \cellcolor{white}\scalebox{1.5}{\ref{plot:line23}} & \cellcolor{white}\scalebox{1.5}{\ref{plot:line22}} &  \cellcolor{white}\scalebox{1.5}{\ref{plot:line21}}   \\
\hline 
\scriptsize \cellcolor{white}$2^{18}$ & \cellcolor{white}\scalebox{1.5}{\ref{plot:line33}} & \cellcolor{white}\scalebox{1.5}{\ref{plot:line32}} &  \cellcolor{white}\scalebox{1.5}{\ref{plot:line31}}   \\
\hline 
\end{tabular} 	}
};

	\end{tikzpicture}

%\end{document}
}
	%\vspace{-9pt}
	\caption{Runtime depending on tier count, number of MACs and workload parameter $K$ ($M = 64, N = 147$).}
	\label{fig:xIsEll}
	\vspace{8pt}
	\scalebox{.81}{
	%\documentclass[border=2mm]{standalone}
%\usepackage[dvipsnames]{xcolor}
%\usepackage{pgfplots}
%\pgfplotsset{compat=1.8}
%\usepackage{amssymb}
%\usepackage{units}
%\usetikzlibrary{decorations.pathreplacing, decorations, positioning, calc, patterns, arrows, math, shadings, matrix}
%\input{defs.tex}

%\begin{document}
	\begin{tikzpicture}
		\begin{axis}[
		ymode=log,
		axis line style={lightgray},
		grid,
		ymax = 4,
		ymin = 0.90,
		xmin = 0,
		xmax = 11,
		title style = {font =\small, yshift =-5},
		label style={font=\scriptsize},
		ylabel style = {yshift = -20},
		xlabel style = {yshift = 5},
		ytick={1, 2, 4 },
		yticklabels = {1, 2, 4},
		x tick label style={font=\scriptsize, yshift = 0},
		y tick label style = {font=\scriptsize, xshift = -1},
		legend style = {font = \tiny},
   		legend pos = north west,
   		legend image post style={scale=0.4},
   		legend style={row sep=-4pt},
   		legend columns=3,
   		legend cell align={left},
	%	restrict y to domain*=0:1.5,
%			visualization depends on=rawy\as\rawy,
%			 after end axis/.code={ % Draw line indicating break
%				\draw [ultra thick, white, decoration={snake, amplitude=1pt}, decorate] (rel axis cs:0,1.05) -- (rel axis cs:1,1.05);
%			},
		xlabel = {MAC count budget},
		ylabel = {speedup normalized to 2D},
		xtick= data,
		xticklabels = {$2^{8}$, $2^{9}$, $2^{10}$, $2^{11}$, $2^{12}$, $2^{13}$, $2^{14}$, $2^{15}$, $2^{16}$, $2^{17}$, $2^{18}$, $2^{19}$}, 
		%symbolic x coords={1,2,3,4,5},
		height =5.1cm,
		width = 1.3\linewidth, 
		%		ymin= 0.9
	%	bar width=0.11cm,
		%every axis plot/.append style={fill,draw=none,no markers},
		ytick style={draw=none},
		xtick style={draw=none},clip=false,	]
		
			 \coordinate (rect1) at (axis cs:0.01,1);
		\coordinate (rect2) at (axis cs:10.99,0.901);
		
		\fill[red!10] (rect1) rectangle(rect2);
		
		\addplot[myMarkerPlot, col1, mark = square*] table[x=mac,y=k1024n64, col sep=comma] {data/MACCount/plot2resultsSpeedup.csv};\label{plot:2:line3}
		\addplot[myMarkerPlot, col1, mark = triangle*] table[x=mac,y=k8192n64, col sep=comma] {data/MACCount/plot2resultsSpeedup.csv};\label{plot:2:line2}
		\addplot[myMarkerPlot, col1, mark = *] table[x=mac,y=k16364n64, col sep=comma] {data/MACCount/plot2resultsSpeedup.csv};\label{plot:2:line1}

		\addplot[myMarkerPlot, col2, mark = square*] table[x=mac,y=k1024n512, col sep=comma] {data/MACCount/plot2resultsSpeedup.csv};\label{plot:2:line23}
		\addplot[myMarkerPlot, col2, mark = triangle*] table[x=mac,y=k8192n512, col sep=comma] {data/MACCount/plot2resultsSpeedup.csv};\label{plot:2:line22}
		\addplot[myMarkerPlot, col2, mark = *] table[x=mac,y=k16364n512, col sep=comma] {data/MACCount/plot2resultsSpeedup.csv};\label{plot:2:line21}

		\addplot[myMarkerPlot ,col3, mark = square*] table[x=mac,y=k1024n2048, col sep=comma] {data/MACCount/plot2resultsSpeedup.csv};\label{plot:2:line33}
		\addplot[myMarkerPlot, col3, mark = triangle*] table[x=mac,y=k8192n2048, col sep=comma] {data/MACCount/plot2resultsSpeedup.csv};\label{plot:2:line32}
		\addplot[myMarkerPlot, col3, mark = *] table[x=mac,y=k16364n2048, col sep=comma] {data/MACCount/plot2resultsSpeedup.csv};\label{plot:2:line31}

		\addplot [densely dashed, very thick, col1, mark=none] coordinates {(4, 0.9) (4, 4)} node [pos = .65, right] {\scriptsize $\mathcal{N}_{\mathbf{min}}$};
		\addplot [densely dashed, very thick, col2, mark=none] coordinates {(7, 0.9) (7, 4)} node [pos = .57, right] {\scriptsize $\mathcal{N}_{\mathbf{min}}$};
		\addplot [densely dashed, very thick, col3, mark=none] coordinates {(9, 0.9) (9, 4)} node [pos = .45, right] {\scriptsize $\mathcal{N}_{\mathbf{min}}$};
		
		\draw [thick, OrangeRed] ({rel axis cs:0,0}|-{axis cs:2,1}) -- ({rel axis cs:1,0}|-{axis cs:2,1}) node [pos=0.15, above, ForestGreen] {\sffamily \scriptsize $\Uparrow$ better} node [align = left, pos=0.90, below, yshift = 1.5] {\sffamily \scriptsize $\Downarrow$ worse};

	 \coordinate (legend) at (axis description cs:-0.005,1.02);
		\end{axis}

	\node [anchor = north west ]at (legend) {
\begin{tabular}{|c|c|c|c|}

\hline 	
\cellcolor{white}& \multicolumn{3}{c|}{\cellcolor{white}\scriptsize\bfseries K}\\
\scriptsize  \cellcolor{white}\bfseries N &\cellcolor{white} \scriptsize $2^{10}$ & \cellcolor{white}\scriptsize $2^{13}$ & \cellcolor{white}\scriptsize $2^{14}$\\
\hline 
\cellcolor{white}\scriptsize $2^{6}$ & \cellcolor{white}\scalebox{1.5}{\ref{plot:2:line3}} & \cellcolor{white}\scalebox{1.5}{\ref{plot:2:line2}} & \cellcolor{white}\scalebox{1.5}{\ref{plot:2:line1}}   \\
\hline 
\cellcolor{white}\scriptsize $2^{9}$ & \cellcolor{white}\scalebox{1.5}{\ref{plot:2:line23}} &\cellcolor{white}\scalebox{1.5}{\ref{plot:2:line22}} & \cellcolor{white}\scalebox{1.5}{\ref{plot:2:line21}}   \\
\hline 
\cellcolor{white}\scriptsize $2^{11}$ & \cellcolor{white}\scalebox{1.5}{\ref{plot:2:line33}} & \cellcolor{white}\scalebox{1.5}{\ref{plot:2:line32}} & \cellcolor{white}\scalebox{1.5}{\ref{plot:2:line31}}   \\
\hline 
\end{tabular} 	
};

	\end{tikzpicture}

%\end{document}
}
	%\vspace{-20pt}
	\caption{Runtime depending budget of MACs and workload parameters $N$ and $K$ (4 tiers, $M = 64$).}
	\label{fig:macCount}
\end{figure}

We compare the performance of 3D and 2D using Eq.~\ref{eq:2d} and Eq.~\ref{eq:model}. We assume an identical number of MACs that are evenly split up among tiers. (Eq.~\ref{eq:2d}  holds with $\mathcal{N} = RC$ and Eq.~\ref{eq:model} holds with $\lfloor \mathcal{N}/\ell \rfloor=  R'C'$.) We round down to avoid resource over-provision. 

%In 3D, the workload profits from increased spatial parallelism, but each tier provides less computational power than a single large 2D tier. 
%The performance of a 3D array depends on the workload parameters ($M$, $N$, and $K$); the number of tiers ($\ell$); and the number of MACs ($\mathcal{N}$). We will show that 3D can achieve better performance vs.\ 2D if these parameters are suitable. %In the next paragraphs, we will discuss the influence of these parameters on runtime to show the design implications for 3D systolic arrays.

\subsubsection{Workload parameters}

\emph{The influence of $K$}, the inner dimension of the matrix-matrix product, is shown in Fig.~\ref{fig:xIsEll}. It depicts the speedup of a 3D-accelerator normalized to its 2D-counterpart with same MAC count (y-axis) depending on the tier count (x-axis). There are different curves for varying number of MACs (same color) and varying parameter $K$ (same shape). For each curve, $M$ and $N$ are fixed. The workloads are taken from a language recognition network (Resnet50). 

The performance of the 3D array improves for larger $K$ and a fixed MAC count. 3D is not advantageous for a small $K$ and a small MAC count (e.g., $K=255$ and $2^{12}$ MACs), as of worse performance than in 2D (green plots). If $K$ is large, the 3D array yields a significant speedup. In best case, we see a speedup of up to 1.93$\times$ for 2 tiers and up to 9.16$\times$ for 12 tiers vs. 2D.

%on performance is shown in Fig.~\ref{fig:xIsEll}: The speedup normalized to a 2D array is plotted  depending on tier count, number of MACs and the workload parameter $K$, while $M$ and $N$ are fixed. The workload parameters are taken from a language recognition network (Resnet50). Markers with the same shape denote same values of $K$. For all MAC counts, the performance of the 3D array improves for larger $K$. If $K$ and the MAC count are too small (e.g. $K=255$ and 4096 MACs), the 3D array will perform worse than a 2D array (cf. green plots). If $K$ is large enough, the influence of the MAC count and the number of tiers will be smaller (cf. purple plots). In best case, we see a speedup of up to 1.93x for 2 tiers and up to 9.16x for 12 tiers vs. 2D. To summarize, 3D integration improves performance for workloads with large $K$.

\emph{The influence of $M$/$N$}, the outer dimensions of the matrix product are shown in Fig.~\ref{fig:macCount}. It depicts speedup of a 3D-accelerator normalized to its 2D-counterpart with same MAC count (y-axis) depending on a given budget of MACs, i.e., processing power (x-axis). We set the number of tiers to 4. There are different curves for varying $K$ (same color) and varying $N$ (same shape). The influence of $M$ and $N$ is symmetrical, so we only vary $N$, while $M$ is constant. 

The parameter $N$ and $M$ determine a threshold $\mathcal{N}_{min}$ for a minimal MAC count required to gain a performance benefit from 3D; the threshold is marked with a dashed line. The threshold is given by $\mathcal{N}_{min} > MN$ (This was evaluated for reasonable tier counts $\leq$16, although not shown). We achieve a maximum speedup of 3.13$\times$ for the given parameter sets.

To summarize, the workload analysis shows that 3D arrays provide a large performance benefit vs.\ 2D for workloads with large $K$ and relatively small $M$ and $N$. This is often the case, e.g., in language recognition models. The minimal MAC count to gain a speedup is given by $MN$.

\subsubsection{Architectural parameters}

The MACs count must be high to unleash 3D-integration, as shown In Fig.~\ref{fig:xIsEll}. For instance, for the workload with $K = 255$ yields a 51\% performance loss for $2^{12}$ MACs but up to 9.16$\times$ speedup for $2^{18}$ MACs compared against 2D. 

Fig.~\ref{fig:macCount} shows the required MAC budget for which 3D provides speedup as a dotted vertical line. For MAC budgets larger than the threshold, there is a continuous performance improvement until saturation, for which provision of additional computational power does not make sense.

Fig.~\ref{fig:xIsEll} shows the influence of the tier count. More tiers continue the trend for a given workload, i.e., if the workload yields better 3D performance, more tiers further will improve the performance and vice versa. Local minima for different tier counts are artifacts of quantization. 

Any speedup of 3D is reduced for very large tier counts as the reduction of partial sums overtakes the time of partial sum generation, cf. Eq.~\ref{eq:model}. As the tier count is limited by production, we do not further discuss this. 

\begin{figure}
\centering
\includegraphics[width = .85\linewidth]{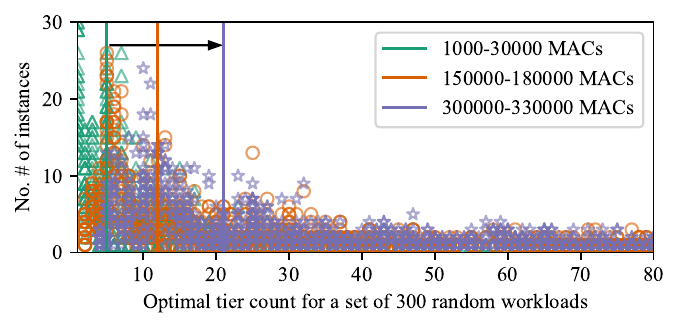}
%\vspace{-12pt}
\caption{Scatter plot of optimal tier for different MAC count. Median marked as vertical line.}
\label{fig:optimaltiers}
\end{figure}

Fig.~\ref{fig:optimaltiers} show the combined influence of the tier count and MAC count. It is a scatter plot of the optimal tier count for a set of 300 random workloads based on Resnet50 parameters. The data are plotted for three MAC budgets resulting in a tail-heavy and shifted right distribution of the optimal tier count for larger MAC budgets. A vertical line shows the median of each distribution; the shift is highlighted by the black arrow. We conclude a trend that 3D arrays with larger MAC counts profit from larger tier counts. 

To summarize, our architectural analysis shows that 3D is a viable choice for large system. Small devices with less than 4096 MACs will require other innovations to gain performance from 3D. 3D currently only targets high-performance systems such as servers due to high production costs so that our finding does not impede the practical application of 3D for DNN-accelerators. A high tier count would be favorable, although this is not realizable with current 3D manufacturing. 

\vspace{\subsectionRemoveSpace}
\subsection{Power}
We compare the power of a 3D-IC with TSVs or MIVs against a 2D-IC. TSVs have a very high capacitance of about 10fF \cite{Song.2013}, while MIV only have about 0.2fF capacitance \cite{Samal.2016}. 

We found that a static power analysis is insufficient. The reason lies in the special dataflow of a 3D-array, in which the horizontal links are heavily utilized while the vertical links are only used for partial sum accumulation. Hence, the switching activities of horizontal and vertical links vary.  

We conduct post-synthesis power analysis  for a 3-layer IC in \unit[15]{nm} node with an example workload of $M, N$=$128$ and $K$=$300$ using Synopsys\textsuperscript{\textregistered} PrimeTime PX. 

The power consumption of an array with 16384 MACs per layer is shown in Tab.~\ref{tab:staticpower} (excluding power from data transmission from memory), along with the difference in power consumption vs.\ a 2D-IC with a similar total number of MACs (49284 MACs). We find that TSV-based 3D-IC requires  5.39\% less power than a 2D-IC and a MIV-based 3D-IC requires 2.21\% less power. As expected, MIVs are more frugal than TSVs. 

3D-ICs do draw less power than 2D-ICs because of the special properties of the dataflow. This demonstrates the relevance of dynamic power analysis for 3D systolic arrays.

\begin{table}
	\caption{Power of a 3D array with 16384 MACs and 3 layers vs. a 2D array with similar number of 49284 MACs.}
	\label{tab:staticpower}	
	\centering
	%\vspace{-6pt}
	{\scriptsize \begin{tabular}{l|r|r|r|r}
		\toprule
		& Total Power & $\Delta$ & Peak Power & $\Delta$ \\ 
		\midrule
		2D & 		\unit[6.61]{W}& 	--- & 		\unit[14.99]{W} & 	---\\ 		
		3D TSV &  	\unit[6.39]{W}& 	-5.4\% &	\unit[14.41]{W} & 	-5.9\%\\ 
		3D MIV &  	\unit[6.26]{W}& 	-2.2\% & 	\unit[14.14]{W} & 	-2.1\%\\ 
		\bottomrule
	\end{tabular} }
\end{table}

\vspace{\subsectionRemoveSpace}
\subsection{Thermal Performance}
%In general, thermal performance is one of the most urgent issues of 3D integration \cite{Perry.2019}. 
As thermal performance is one of the most urgent issues of 3D integration \cite{Perry.2019}, we conduct a thermal analysis with HotSpot 6.0 \cite{Stan.2015}. We chose a three-layer 3D-IC with 4096, 16384 and 65536 MACs per layer and a workload of $M, N = 128$ and $K = 300$. The respective 2D-IC has as 12321, 49284 and 197136 MACs, which is approximately the MAC count of the 3D case. 

The results are shown in Fig.~\ref{fig:temp} as a boxplot. For 3D, we split the data into the layer near the heatsink (\emph{bottom}) and the rest (\emph{middle}). The temperature variability comes from different switching activities and cooler MACs at the borders of the IC as of their fewer neighbors.

3D and 2D ICs get hotter for larger MAC counts. Furthermore, 3D ICs get hotter than 2D ICs. The TSV-based and the MIV-based 3D-ICs are not exceeding their thermal budget. This is a promising finding for 3D-ICs practically used for DNN-accelerators. 

The MIV-based IC is hotter than the TSV-based IC. This is counter-intuitive due to the difference in parasitics of TSVs and MIVs. The reason lies in the vast number of vertical links. The large TSVs increase area, enhance heat dissipation and reduces the temperature. In a real system, one would apply TSV-saving schemes to improve area and yield, which will increase the temperature above the level of MIV-based ICs.

\begin{figure}
	\centering
	\scalebox{.78}{

\pgfplotsset{
	box plot/.style={
		/pgfplots/.cd,
		black,
		only marks,
		mark=-,
		mark size=\pgfkeysvalueof{/pgfplots/box plot width},
		/pgfplots/error bars/y dir=plus,
		/pgfplots/error bars/y explicit,
		/pgfplots/table/x index=\pgfkeysvalueof{/pgfplots/box plot x index},
	},
	box plot box/.style={
		/pgfplots/error bars/draw error bar/.code 2 args={%
			\draw  ##1 -- ++(\pgfkeysvalueof{/pgfplots/box plot width},0pt) |- ##2 -- ++(-\pgfkeysvalueof{/pgfplots/box plot width},0pt) |- ##1 -- cycle;
		},
		/pgfplots/table/.cd,
		y index=\pgfkeysvalueof{/pgfplots/box plot box top index},
		y error expr={
			\thisrowno{\pgfkeysvalueof{/pgfplots/box plot box bottom index}}
			- \thisrowno{\pgfkeysvalueof{/pgfplots/box plot box top index}}
		},
		/pgfplots/box plot
	},
	box plot top whisker/.style={
		/pgfplots/error bars/draw error bar/.code 2 args={%
			\pgfkeysgetvalue{/pgfplots/error bars/error mark}%
			{\pgfplotserrorbarsmark}%
			\pgfkeysgetvalue{/pgfplots/error bars/error mark options}%
			{\pgfplotserrorbarsmarkopts}%
			\path ##1 -- ##2;
		},
		/pgfplots/table/.cd,
		y index=\pgfkeysvalueof{/pgfplots/box plot whisker top index},
		y error expr={
			\thisrowno{\pgfkeysvalueof{/pgfplots/box plot box top index}}
			- \thisrowno{\pgfkeysvalueof{/pgfplots/box plot whisker top index}}
		},
		/pgfplots/box plot
	},
	box plot bottom whisker/.style={
		/pgfplots/error bars/draw error bar/.code 2 args={%
			\pgfkeysgetvalue{/pgfplots/error bars/error mark}%
			{\pgfplotserrorbarsmark}%
			\pgfkeysgetvalue{/pgfplots/error bars/error mark options}%
			{\pgfplotserrorbarsmarkopts}%
			\path ##1 -- ##2;
		},
		/pgfplots/table/.cd,
		y index=\pgfkeysvalueof{/pgfplots/box plot whisker bottom index},
		y error expr={
			\thisrowno{\pgfkeysvalueof{/pgfplots/box plot box bottom index}}
			- \thisrowno{\pgfkeysvalueof{/pgfplots/box plot whisker bottom index}}
		},
		/pgfplots/box plot
	},
	box plot median/.style={
		/pgfplots/box plot,
		/pgfplots/table/y index=\pgfkeysvalueof{/pgfplots/box plot median index}
	},
	box plot width/.initial=1em,
	box plot x index/.initial=0,
	box plot median index/.initial=1,
	box plot box top index/.initial=2,
	box plot box bottom index/.initial=3,
	box plot whisker top index/.initial=4,
	box plot whisker bottom index/.initial=5,
}

\newcommand{\boxplot}[2][]{
	\addplot [box plot median,#1] table {#2};
	\addplot [forget plot, box plot box,#1] table {#2};
	\addplot [forget plot, box plot top whisker,#1] table {#2};
	\addplot [forget plot, box plot bottom whisker,#1] table {#2};
}

%\begin{document}
	\begin{tikzpicture}
	\begin{axis} [box plot width=1.5mm, 
	xmin = -6.5,
	xmax = 10.5,
height =5.6cm,
width = 1.3\linewidth, 
		axis line style={lightgray},
		ymajorgrids,
	ylabel = {temperature [K]},
	label style={font=\scriptsize},
	ylabel style = {yshift = -15},
	xlabel style = {yshift = 5},
	x tick label style={font=\scriptsize, yshift = 0, align = center},
	xtick={-6, -5, -4, -3, -2, 0,1,2,3,4, 6,7,8,9,10},%, 12, 13,14, 15, 16},
	xticklabels = {2D, 3D\\TSV\\\tiny bottom, 3D\\TSV\\\tiny middle, 3D\\MIV\\\tiny bottom, 3D\\MIV\\\tiny middle,2D, 3D\\TSV\\\tiny bottom, 3D\\TSV\\\tiny middle,
	 3D\\MIV\\\tiny bottom, 3D\\MIV\\\tiny middle, 2D, 3D\\TSV\\\tiny bottom, 3D\\TSV\\\tiny middle, 3D\\MIV\\\tiny bottom, 3D\\MIV\\\tiny middle,  },
	ytick = {325,330,335,340,345},
	y tick label style = {font=\scriptsize},
	]
	\boxplot [forget plot,  col3] {data/temp/temps0.dat}
	\boxplot [forget plot,  col1] {data/temp/temps1.dat}
	\boxplot [forget plot, thick, col2] {data/temp/temps2.dat}
%	\boxplot [forget plot, col3] {data/temp/temps3.dat}
	\coordinate (legend) at (axis description cs:-0.005,1.03);
	\end{axis}
	
		\node [anchor = north west ]at (legend) {
		\begin{tabular}{|c|c|}
		
		\hline 	
		\scriptsize  \cellcolor{white}\bfseries tiers $\times$ MAC &\cellcolor{white} \\
		\hline 
		\cellcolor{white}\scriptsize $3\times2^{12}$  & \cellcolor{white}\begin{tikzpicture}\fill[col3] (0,0) rectangle(.15,.15);\end{tikzpicture}\\
		\cellcolor{white}\scriptsize $3\times2^{14}$  & \cellcolor{white}\begin{tikzpicture}\fill[col1] (0,0) rectangle(.15,.15);\end{tikzpicture}\\
		\cellcolor{white}\scriptsize $3\times2^{16}$  & \cellcolor{white}\begin{tikzpicture}\fill[col2] (0,0) rectangle(.15,.15);\end{tikzpicture}\\
%		\cellcolor{white}\tiny 3 layers, 262144 & \cellcolor{white}\begin{tikzpicture}\fill[col3] (0,0) rectangle(.1,.1);\end{tikzpicture}\\
		\hline
		\end{tabular} 	
	};
	
	\end{tikzpicture}
%\end{document}
}
		%\vspace{-12pt}
	\caption{Boxplot of the temperature of different 2D and 3D arrays for an example workload with $M, N = 128$, $K = 300$.}
	\label{fig:temp}
	\vspace{8pt}
		\scalebox{.9}{
		%\documentclass[border=2mm]{standalone}
%\usepackage[dvipsnames]{xcolor}
%\usepackage{pgfplots}
%\pgfplotsset{compat=1.8}
%\usepackage{amssymb}
%\usepackage{units}
%\usetikzlibrary{decorations.pathreplacing, decorations, positioning, calc, patterns, arrows, math, shadings, matrix}
%\input{defs.tex}

%\begin{document}
	\begin{tikzpicture}
		\begin{axis}[
		ymode=log,
		axis line style={lightgray},
		grid,
		ymax = 8,
		ymin = .25,
		xmin = 2,
		xmax = 12,
		title style = {font =\small, yshift =-5},
		label style={font=\scriptsize},
		ylabel style = {yshift = -17},
		xlabel style = {yshift = 5},
		x tick label style={font=\scriptsize, yshift = 0},
		ytick={.25, .5, 1, 2, 4 , 8},
		yticklabels = {.25, .5, 1, 2, 4, 8},
		y tick label style = {font=\scriptsize},
		legend style = {font = \tiny},
   		legend pos = north west,
   		legend image post style={scale=0.4},
   		legend style={row sep=-4pt},
   		legend columns=3,
   		legend cell align={left},
	%	restrict y to domain*=0:1.5,
%			visualization depends on=rawy\as\rawy,
%			 after end axis/.code={ % Draw line indicating break
%				\draw [ultra thick, white, decoration={snake, amplitude=1pt}, decorate] (rel axis cs:0,1.05) -- (rel axis cs:1,1.05);
%			},
		xlabel = {tier count},
		ylabel = {Performance per area},
		xtick= data,
%		xticklabels = {256, 512, 1024, 2048, 4096, 8192, 16384, 32768, 65536, 131072, 262144, 524288}, 
		%symbolic x coords={1,2,3,4,5},
		height =4.3cm,
		width = .8\linewidth, 
		%		ymin= 0.9
	%	bar width=0.11cm,
		%every axis plot/.append style={fill,draw=none,no markers},
		ytick style={draw=none},
		xtick style={draw=none},clip=false,	]
		
		\coordinate (rect1) at (axis cs:2.01,0.251);
		\coordinate (rect2) at (axis cs:11.99,0.99);
		\fill[red!10] (rect1) rectangle(rect2);
%		
%		\coordinate (rect3) at (axis cs:2.01,1.01);
%		\coordinate (rect4) at (axis cs:11.99,7.99);
%		\fill[green!10] (rect3) rectangle(rect4);

		\addplot[myMarkerPlot, col2, mark = square*] table[x=l,y=macs4096tsv, col sep=comma] {data/Area/areaPerf.csv};\label{plot:area:line3}
		\addplot[myMarkerPlot, col2, mark = triangle*] table[x=l,y=macs32768tsv, col sep=comma] {data/Area/areaPerf.csv};\label{plot:area:line2}
		\addplot[myMarkerPlot, col2, mark = *] table[x=l,y=macs262144tsv, col sep=comma] {data/Area/areaPerf.csv};\label{plot:area:line1}

\addplot[myMarkerPlot, col1, mark = square*] table[x=l,y=macs4096miv, col sep=comma] {data/Area/areaPerf.csv};\label{plot:area:line23}
\addplot[myMarkerPlot, col1, mark = triangle*] table[x=l,y=macs32768miv, col sep=comma] {data/Area/areaPerf.csv};\label{plot:area:line22}
\addplot[myMarkerPlot, col1, mark = *] table[x=l,y=macs262144miv, col sep=comma] {data/Area/areaPerf.csv};\label{plot:area:line21}

%\addplot[col3, mark = square*] table[x=l,y=k1024n2048, col sep=comma] {data/Area/areaPerf.csv};\label{plot:2:line33}
%\addplot[col3, mark options={scale=1.5, fill=col3}, mark = triangle*] table[x=l,y=k8192n2048, col sep=comma] {data/Area/areaPerf.csv};\label{plot:2:line32}
%\addplot[col3, mark = *] table[x=l,y=k16364n2048, col sep=comma] {data/Area/areaPerf.csv};\label{plot:2:line31}

		\draw [thick, red] ({rel axis cs:0,0}|-{axis cs:2,1}) -- ({rel axis cs:1,0}|-{axis cs:2,1});
%		node [pos=0.90, above] {\sffamily \scriptsize $\Uparrow$ worse} node [align = left, pos=0.15, below] {\sffamily \scriptsize $\Downarrow$ better};
%		\legend{{TSV, MAC = 4096}, {TSV, MAC = 32768}, {TSV, MAC = 262144},
%			{MIV, MAC = 4096}, {MIV, MAC = 32768}, {MIV, MAC = 262144}};
	 \coordinate (legend) at (axis description cs:1.01,0.5);
		\end{axis}	
%	\matrix [
%	fill = white, 
%	draw,
%	every node/.append style={%
%		inner xsep=0pt,
%		inner ysep=0pt, 
%		draw=lightgray, % just to show node borders
%		text height=10,%\heightof{0},
%		text width=24,
%		align=center
%	},
%	matrix of nodes,
%	anchor=north west,
%	] at (legend) {
%		\tiny MAC\textbackslash K & \tiny \phantom{\textbackslash}12100 & \tiny \phantom{\textbackslash}784 & \tiny \phantom{\textbackslash}255\\
%		\tiny 4096 & \scalebox{1.5}{\ref{plot:2:line1}} & \scalebox{1.5}{\ref{plot:2:line2}} & \scalebox{1.5}{\ref{plot:2:line3}}   \\
%		\tiny 32768 & \scalebox{1.5}{\ref{plot:2:line21}} & \scalebox{1.5}{\ref{plot:2:line22}} & \scalebox{1.5}{\ref{plot:2:line23}}   \\
%		\tiny 262144 & \scalebox{1.5}{\ref{plot:2:line31}} & \scalebox{1.5}{\ref{plot:2:line32}} & \scalebox{1.5}{\ref{plot:2:line33}}   \\
%	};
%
	\node [anchor = west ]at (legend) {
\begin{tabular}{|c|c|c|}

\hline 	
\scriptsize  \cellcolor{white}\bfseries MAC &\cellcolor{white} \bfseries\scriptsize TSV & \cellcolor{white}\bfseries\scriptsize MIV \\
\hline 
\scriptsize $2^{12}$   &\ref{plot:area:line3} & \ref{plot:area:line23}\\
\scriptsize $2^{15}$  &\ref{plot:area:line2} & \ref{plot:area:line22}\\
\scriptsize $2^{18}$ &\ref{plot:area:line1} & \ref{plot:area:line21}\\
\hline
\end{tabular} 	
};

	\end{tikzpicture}

%\end{document}
}
		%\vspace{-12pt}
	\caption{Area-normalized performance for a TSV-/MIV-based 3D array relative to a 2D one ($M$=$64$, $N$=$147$, $K$=$12100$).}
	\label{fig:PerfArea}
\end{figure}

\vspace{\subsectionRemoveSpace}
\subsection{Area}\label{sec:area}

We implement the 2D and 3D arrays in a \unit[15]{nm} node with 8b inputs and 16b outputs for \unit[1]{GHz} clock frequency. We take TSV area plus keep-out-zone (KOZ) from \cite{Song.2013} and MIV area from \cite{Chang.2016}. 

The TSV-based 3D-IC is larger than the 2D array from additional area for logic, TSVs and KOZs. Monolithic integration only adds a few percent overhead vs. 2D, as no KOZs are required.

We plot the runtime per chip area to evaluate the area-impled trade-offs. This is plotted in Fig.~\ref{fig:PerfArea} normalized against 2D for different tier counts for one exemplary given workload from Resnet50. Based on our previous discussion for runtime, we chose a workload that yields a performance benefit for 3D ($M$=$64$, $N$=$147$, $K$=$12100$). The results are shown in Fig.~\ref{fig:PerfArea} for a TSV-based (orange) and for a MIV-based (purple) 3D-IC.

For 4096 and 32768 MACs, the performance per area of the 3D-IC is worse by up to 75\% than the 2D-IC. For 266144 MACs, the area per performance is improved for more than 4 layers by 1.27$\times$ to 2.83$\times$ (cf. \ref{plot:area:line1}). This finding underlines again that 3D integration is useful for large MAC counts. 

We took a worst-case approach to the TSV count, as we provide a dedicated TSV array connecting each pair of MACs. If we apply TSV-reduction architectures (cf. Sec.~\ref{sec:3d-arrays:arch}), TSV-based 3D-ICs will come off better. 

MIV-based 3D-IC enable a better performance per area than TSVs: While the performance per area for 4096 MACs is similar to 2D-ICs, MIV-3D-ICs improve performance per area by up to 7.9$\times$ for larger MAC counts. The general trend is that higher MAC counts and number of tiers improve the performance advantage of 3D vs.\ 2D. 

Two tiers with face-to-face bonding can be manufactured at time of writing this paper. For this, 3D integration allows for 1.19$\times$ to 1.97$\times$ better performance per area.

\vspace{\sectionRemoveSpace}
\section{Conclusion}\label{sec:conclusion}

In this paper, the implications of 3D-ICs for DNN-accelerators on their architecture, dataflow and design are analyzed. 3D-integration allows to add an additional level of spatial parallelism that is otherwise executed in the time domain for a 2D system. 
We choose a systolic-array based architecture and propose a 3D-implementation.
%After discussing memory integration, 
We describe a suitable dataflow \emph{distributed output stationary} that fully utilizes the capability of 3D and is not equivalent to existing data mappings for 2D. 
Using an analytical performance model and an RTL implementation for the 3D-array, we conduct an in-depth analysis about design implications in computational and thermal performance, area and power.
Our analysis depicts that 3D-implementation enables performance improvements for DNN workloads.
We identify a threshold for required computational performance to fully gain a speedup for 3D. The speedup is almost an order of magnitude (up to 9.14x) vs. 2D. 
From an architectural perspective, we find that a higher MAC count and more \textit{tiers} improve the performance of 3D, while over-provisioning of computational resources leads to a speedup saturation vs. 2D. We show that thermal performance allows for 3D integration of DNN accelerators both with TSVs and MIVs.
The area of a 3D accelerator is larger than 2D; but even for TSVs the performance per area is superior to 2D as MACs and \textit{tiers} are scaled. 
Monolithic 3D integration naturally offers advantages over TSV-based stacking for area and power. To summarize, we conduct a comprehensive study on 3D-accelerators that is universal in that it abstracts from design details that are not purely related to vertical integration of DNN-accelerators such as memory and TSV-count. 

%\section*{Acknowledgments}
%This work was supported by a fellowship within the IFI programme of the German Academic Exchange Service (DAAD).

\bibliographystyle{IEEEtran}
\vspace{-1ex}
\renewcommand\refname{}
\section*{References}
\vspace{-4ex}
\bibliography{bibliographyShort}
%\bibliographystyle{ACM-Reference-Format}
%\bibliography{bibEtAl}

\end{document}